\documentclass[twoside]{article}

\usepackage{graphicx}

\usepackage{cmpj2e}

\title[Spin current in frustrated chains]%
{Field-controlled spin current in frustrated spin chains}

\author{A.K. Kolezhuk\refaddr{label1,label2},
        I.P. McCulloch\refaddr{label3}}
\addresses{
\addr{label1} Institut f\"ur Theoretische Physik C, RWTH Aachen University, 52056 Aachen,
  Germany
\addr{label2} Institute of
Magnetism, National Academy of Sciences and Ministry of Education,
03142 Kiev, Ukraine
\addr{label3} Department of Physics, University of Queensland,
Brisbane QLD 4072, Australia
}

\begin{document}

\maketitle

\begin{abstract}
We study  states with spontaneous spin current,
emerging in frustrated antiferromagnetic spin-$S$ chains subject to a strong external
magnetic field. As  a numerical tool, we use  a non-Abelian symmetry realization of the
density matrix renormalization group.
The field dependence of the
order parameter and the critical exponents are presented for zigzag chains with $S=1/2$,
$1$, $3/2$, and $2$. 
\pacs 75.10.Pq, 75.40.Cx, 75.40.Mg
\end{abstract}

\section{Introduction}

In usual magnets, spin degrees of freedom order ferro- or antiferromagnetically,
while those two basic orders may be viewed as special cases of a general spin
density wave order parameter. It had been known for a long time
\cite{Villain78,AndreevGrishchuk84} that other ``exotic'' spin orderings with a
vanishing expectation value of the spin density are theoretically possible, but
only recently such orders have been observed in numerical studies of realistic
models \cite{Kaburagi+99,Hikihara+01,Shannon+06,McCulloch+08,Hikihara+08}. One prominent
example of unconventional order parameter is the spin current (the so-called
vector chirality \cite{Villain78}, or $p$-type nematic \cite{AndreevGrishchuk84}
order parameter). In a spin chain, the vector chirality operator
$\vec{\kappa}(n)\equiv \vec{\kappa}_{n, n+1}$ can be defined as a vector product of two adjacent spins:
\begin{equation}
\label{kappa-def}
\vec{\kappa}_{n, n+1}= (\vec{S}_{n}\times \vec{S}_{n+1}),
\end{equation}
where $n$ labels the lattice sites along the chain. Obviously, the vector chiral
order is nonzero in a state which has a classical helical magnetic order, and the
projection of $\vec{\kappa}$ onto the helix axis distinguishes between the left
and right spirals.  For low-dimensional magnets, such unconventional order
parameters may become dominating, since quantum fluctuations have a tendency to
destroy the usual magnetic order. A state with vector chiral order may be viewed
as the result of phase fluctuations destroying the helical order, but still
keeping the preferred sense of rotation contained in a classical spiral
\cite{ChandraColeman91}. The interpretation of $\vec{\kappa}$ as the spin
current can be understood from  the equation of motion, which for a Hamiltonian of the form
$\mathcal{H}=\sum_{nm}J_{m}\vec{S}_{n}\cdot\vec{S}_{n+m}$ reads
$\hbar(\partial\vec{S}_{n}/\partial t) = \sum_{m} J_{m} \{\vec{\kappa}_{n-m,n} -\vec{\kappa}_{n,n+m} \}$.
One can show that in the ground state the net spin current should be zero, which
for the above Hamiltonian translates into the condition
$\sum_{m}mJ_{m}\vec{\kappa}_{n,n+m}=0$.

Spin current states have been relatively well studied for anisotropic frustrated
chains \cite{Nersesyan+98,Kaburagi+99,K00,Hikihara+01,Lecheminant+01}.
recently, it has
been shown \cite{KV05,McCulloch+08,Okunishi08,Hikihara+08} that a
strong external magnetic field can be used as a control parameter to drive the spin
current in isotropic frustrated spin chains.  The interest to this topic is
further boosted by the fact that there are currently several
quasi-one-dimensional magnetic materials that are intensely studied as possible
candidates for manifestation of unconventional orders, particularly $ \rm
LiCuVO_4$ \cite{Enderle+05}, $\rm Li_2ZrCuO_4$ \cite{Drechsler+07}, and $\rm
Cu_{2}Cl_{4}$-$\rm H_{8}C_{4}SO_{2}$ ($\rm Sul$-$\rm Cu_{2}Cl_{4}$) \cite{Garlea+09}. We
have previously studied \cite{McCulloch+08} the spin current correlations in
frustrated chains with the spin $S=1/2$ and $1$, and have shown that the
behavior of $S=1/2$ and $S=1$ chains was very different. It was not clear
whether this can be attributed to some general difference in the behavior of
chains with integer and half-integer $S$.  The goal of the present paper is to
answer this question. For that purpose, we numerically study the spin current in
isotropic spin-$S$ zigzag chains with the spin up to $S=2$. It is shown that
chains with $S\geq 1$ behave similarly and exhibit a state with a finite
spontaneous spin current in the entire region of finite magnetization.  In
Sect.\ \ref{sec:theo} we give a brief overview of the theory of spin current
correlations, Sect.\ \ref{sec:dmrg} presents the numerical results, and
Sect.\ \ref{sec:summary} contains a short summary.


\section{Frustrated chain in magnetic field: spin current correlations}  
\label{sec:theo}

We consider  a frustrated antiferromagnetic spin chain described by the following Hamiltonian:
\begin{equation}
\label{S-ham}
{\mathcal H}=J_{1}\sum_{n} \vec{S}_{n}\cdot \vec{S}_{n+1} 
+J_{2}\sum_{n}\vec{S}_{n}\cdot \vec{S}_{n+2} -H \sum_{n}S_{n}^{z}
\end{equation}
where $\vec{S}_n$ are spin-$S$ operators at the $n$-th site,  
$J_{1}>0$ and $J_{2}>0$ are the nearest and next-nearest neighbor exchange
constants, respectively, and $H$ is the external magnetic field, assumed to
be applied along the $z$ axis.
At zero field, if the frustration parameter $\alpha\equiv J_{2}/J_{1}$ is large
enough, the ground state is generally expected to have a finite spectral gap $\Delta$, both for
integer and half-integer $S$. 
When the applied field exceeds the critical value $H_{c}=\Delta$,
the system acquires finite magnetization. Further increase of the field beyond
the saturation field value
$H_{s}$ brings the system into a fully polarized state. 

 Let us first assume that
$\alpha\gg 1$, so that one deals with the limit of two weakly coupled spin-$S$
subchains.
The behavior of spin chains  in magnetic field have been extensively
studied for $S=1/2$ \cite{Bogoliubov+86,AffleckOshikawa99,Furusaki+03-04} and $S=1$ 
\cite{KonikFendley02,CamposVenuti+02,Sato06}. 
We will assume that for $H_{c}<H < H_{s}$
the low-energy physics of a single chain is well described
in terms of the effective Tomonaga-Luttinger liquid (TLL) theory with the
following Hamiltonian:  
\begin{equation}
\label{SpinChainBosHam}
{\cal H}_{TL}[\theta,\varphi] =  \frac{v}{2}\int dx \, \Big\{\frac{1}{K}(\partial_x \varphi)^{2} 
+ K (\partial_x \theta)^{2}\Big\}.
\end{equation}
Here  $K$ is the so-called TLL parameter,  $v\propto J_{2}$ is the Fermi
velocity, $\phi$ is the compact bosonic field
($\varphi \equiv \varphi+\sqrt{\pi}$),
and  $\theta$ is its dual satisfying the commutation
relations $[\varphi(x),\theta(y)] = i\Theta (y-x)$, where $\Theta(x)$ is the
Heaviside function.

We assume further, by the analogy with the $S=1/2$ and $S=1$ case
\cite{Bogoliubov+86,KonikFendley02,Sato06}, 
that in the continuum limit the most relevant low-energy part of  spin operators can be represented  as
\begin{eqnarray}
\label{spinops}
S^{z}_{a}(x_{a})&=&MS+\frac{2}{\sqrt{\pi}} \partial_x \varphi_{a}(x_{a})
+ A_{3} \sin \big[k_{F} x_{a}+ \sqrt{4\pi }
\varphi_{a}(x_{a})\big]   \\
S^{+}_{a}(x_{a})& =& e^{i \pi x/2} e^{i\sqrt{\pi}\theta_{a}(x_{a})}
\big\{ A_{1} +A_{2}\sin{\big[k_{F} x_{a}+ \sqrt{4\pi }\varphi_{a}(x_{a})\big]}
\big\} ,\nonumber
\end{eqnarray}
Here $a=1,2$ labels the two subchains, the space coordinate $x$ is defined at
the middle of every bond along
the original zigzag chain,  $x_{a}=x+a-3/2$ define the  sites
of the subchains, $MS$ is the ground state magnetization
per spin, $k_{F}$ has a meaning of the Fermi momentum and is connected to $M$
(e.g., $k_{F}=\pi(1+M)/2$ for $S=1/2$), $A_{i}$ are nonuniversal amplitudes. All 
 parameters ($K$, $v$, $M$, $A_{i}$) should be
understood as functions of the field $H$ which generally have to be
extracted by comparing the numerical results with those following from the TLL description
\cite{Bogoliubov+86,AffleckOshikawa99,Furusaki+03-04,KonikFendley02,CamposVenuti+02,Fath03}.

Introducing symmetric and antisymmetric combinations of the bosonic fields
$\varphi_{\pm}=(\varphi_1\pm \varphi_2)/\sqrt{2}$,
$\theta_{\pm}= (\theta_1 \pm \theta_2)/\sqrt{2}$,
one obtains two coupled TLLs, and the nature of the leading interaction term
strongly depends on the value of the TLL parameter $K$. 
The longitudinal ($S^{z}S^{z}$) part of the zigzag exchange leads
to a splitting of the
TLL parameter values for the symmetric  and antisymmetric
 sectors, which for $\alpha\gg 1$ are found as
$ K_{\pm}\approx K\big[1\pm 2K/(\pi v \alpha))\big]^{-1/2}$, $ v_{\pm}\approx v\big[1\pm 2K/(\pi v \alpha))\big]^{1/2}$.
At the same time, the $S^{z}S^{z}$ part of the interaction gives rise to the interaction term proportional to 
$\cos\big[ \sqrt{8\pi }\varphi_{-}
-\pi M\big]$ whose  scaling dimension is $2K_{-}$ and which favors
$n$-type spin-nematic (or XY2 in the
nomenclature of Ref.\ \cite{Schulz86}) correlations \cite{Vekua+07}. The transversal part of the
zigzag exchange yields the so-called ``twist term'' $\sin\big(\sqrt{2\pi}
\theta_{-}\big)(\partial_{x}\theta_{+})$ that has a nonzero conformal spin and
the formal scaling dimension $1+1/(2K_{-})$.  The twist term favors states with
spontaneous spin current \cite{Nersesyan+98}.
If $K<1$, as is the case for $S=1/2$, the two interaction terms compete,
which leads to the Ising-type transition between the nematic and spin-current
state \cite{KV05}. For $S=1$ it is known that $K>1$
\cite{KonikFendley02,CamposVenuti+02,Fath03}, thus the nematic-favoring term is
irrelevant and the spin-current favoring term dominates. For higher spins, to
our knowledge, there is no numerical data on the behavior of the TLL
parameter $K$. However, since the analysis based on a mapping to the
nonlinear $\sigma$-model \cite{KonikFendley02} can be extended to
higher $S$, one may expect that for all spin-$S$ chains with $S\geq
1$, the TLL parameter $K$  is greater than $1$, at least in the region of small (but
finite) magnetization $M$. 

Inside the phase with a spontaneous spin current, the
twist term is the most relevant perturbation, and the mean-field treatment along
the lines of Ref.\ \cite{Nersesyan+98} leads to the conclusion\cite{KV05} that
both $\langle \partial_{x}\theta_{+} \rangle$ and $\langle \sin\big(\sqrt{2\pi}
\theta_{-}\big) \rangle$ should be nonzero, and the antisymmetric sector should
be massive.
Expressing the spin current  $\vec{\kappa}(n)$ as defined in (\ref{kappa-def}) through the
bosonic fields, one obtains \cite{McCulloch+08}
\begin{eqnarray} 
\label{kappa-bos}
\kappa^{z}(x)&=& \sin\big(\sqrt{2\pi} \theta_{-}\big) \Big\{ A_{1}^{2}-
(\pi A_{1}/2)^{2}(\partial_{x}\theta_{+})^{2} 
+(A_{2}^{2}/2)\cos\big(\sqrt{8\pi}\varphi_{+} +2k_{F} x\big) \Big\},\nonumber\\
\kappa^{+}(x) &=& 2A_{1}M \sin\big(\sqrt{\pi/2} \theta_{-}\big) 
\exp\big\{ i\big[ \pi x/2 +\sqrt{\pi/2}\,\theta_{+} \big]\big\} .
\end{eqnarray}
where we have omitted the contributions from massive fields 
and operators with higher scaling
dimensions. 
At large distances $x\gg \xi$, where $\xi$ is the largest correlation length determined by the smallest gap in the
neglected massive fields,
the asymptotics of the leading spin-current correlations takes the form
\begin{eqnarray} 
\label{kappa-corr} 
&& \langle \kappa^{z}(x)\kappa^{z}(0)\rangle \to \kappa_{0}^{2}\big(1+ C_{1}x^{-4}
+C_{2}\cos(2k_{F}x) x^{-4K+}\big) ,\nonumber\\
&& \langle \kappa^{+}(x)\kappa^{-}(0)\rangle \to A_{1}^{2} 
M^{2} x^{-1/(4K_{+})} \exp\{i Q x\},
\end{eqnarray}
where the incommensurate wave vector $Q$ is given by
$Q=\pi/2 +\sqrt{\pi/2}\langle \partial_{x}\theta_{+}\rangle$ and $C_{i}$ are
some numbers.

The asymptotics (\ref{kappa-corr}), derived by means of bosonization, is
expected to be valid in the limit $\alpha\gg1$ and for $H$
sufficiently far from any of 
the critical fields $H_{c}$, $H_{s}$ (close to $H_{c}$ or $H_{s}$ the
bosonization approach becomes inapplicable since the effective bandwidth
goes to zero and the interactions cease to be small compared to the
bandwidth). 
In a close vicinity of the saturation field $H_{s}$, large-$S$ analysis 
\cite{KV05} maps the system to a two-component dilute Bose gas
with repulsive interaction, and the external magnetic field drives the
system into a condensed state,
playing the role of the chemical potential. The interspecies repulsion turns out to be strong
enough to satisfy the phase separation condition, so only one of the species
condenses and  the other condensate is depleted. At the end, one deals with the
one-component pseudo-condensate whose physics is again described by a
(one-component) TLL, and the asymptotic form of
the spin current correlators for $H$ close to $H_{s}$ is
given by \cite{KV05}:
\begin{equation} 
\label{kappa-corr-Hs} 
 \langle \kappa^{z}(x)\kappa^{z}(0)\rangle \to  \kappa_{0}^{2} -
C/x^{2},\quad
 \langle \kappa^{+}(x)\kappa^{-}(0)\rangle \to C'x^{-1/(2K')}  e^{i Q' x}.
\end{equation}
where $\kappa_{0}^{2}\propto (H_{s}-H)$, $C'\propto
(H_{s}-H)^{1/2-1/(4K')}$, $K'>1$ is another TLL parameter
(different from $K$ considered above), and the wave vector $Q'$  is
just the
pitch of the classical helix, $Q'=\pm(\pi -\arccos(1/4\alpha))$.

\section{Results of numerical study}
\label{sec:dmrg}

\begin{figure}[tb]
\centerline{\includegraphics[width=0.65\textwidth]{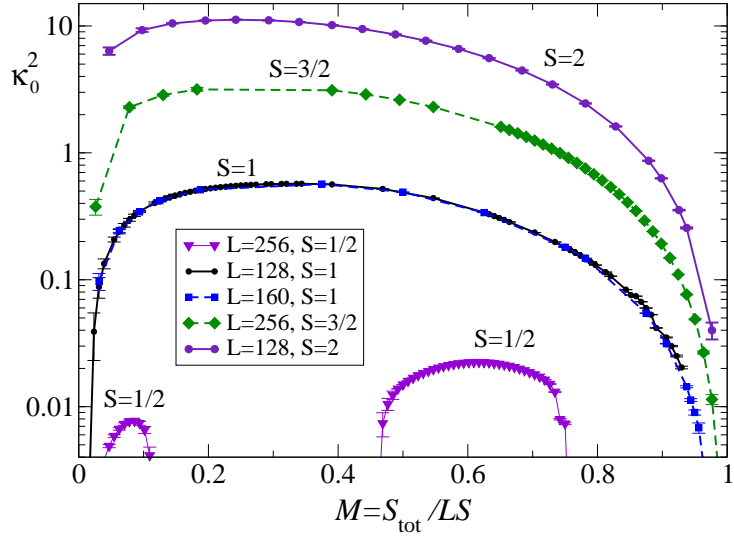}}
\caption{Square of the ground state nearest-neighbor spin current $\kappa_{0}$
  (longitudinal vector chirality) as a function of the magnetization per spin
  $M=S_{\rm tot}/LS$, for isotropic spin-$S$ chains with the spin $S=1/2$, $1$,
  $3/2$ and $2$.  The error bars shown correspond to the uncertainties of the
  fit.}
\label{fig:kappa}
\end{figure}

We have studied numerically the frustrated chain model (\ref{S-ham}) for systems with the 
spin $S=1/2$, $1$, $3/2$, and $2$ and the
frustration parameter $\alpha=J_{2}/J_{1}$ was fixed at $\alpha=1$ (so the
chains may be considered as  stripes of the triangular lattice).
We have used the density matrix
renormalization group (DMRG) method \cite{White92,Schollwock-RMP-05} in its
matrix product state formulation \cite{McCulloch07}, making full use of the
non-Abelian $SU(2)$ symmetry \cite{McCullochGulacsi02,McCulloch07}.
Although the external field breaks
$SU(2)$ symmetry, the fact that the Zeeman energy term commutes with the rest of
the Hamiltonian makes it possible to include the effect of the magnetic
field  by calculating the ground state in a sector with
a finite total spin $S_{\rm tot}$. The use of $SU(2)$ symmetry allows one to
reduce substantially
the number of states $m$ which is necessary to describe the
system: essentially, one
treats the multiplet of states of the same total spin as a single representative
state.
A slight disadvantage is that the non-Abelian method allows to compute only reduced
matrix elements (in the sense of the Wigner-Eckart theorem). 
Since the spin current is a vector, the available correlator that is
easily available is the rotationally invariant scalar product $\langle
\vec{\kappa}(n)\cdot \vec{\kappa}(n')\rangle$, which is a mixture of the
longitudinal and transversal spin current correlations. 
This complicates somewhat the
analysis of the  data: from the theoretical analysis in
Sect.\ \ref{sec:theo} it follows that for typical values of $K\sim 1$ the
longitudinal correlations
$\langle \kappa^{z}(x)\kappa^{z}(0)\rangle$  decay to their asymptotic value
much faster than the transversal ones. 
Thus by analyzing the scalar  correlator $\langle
\vec{\kappa}(n)\cdot \vec{\kappa}(n')\rangle$ one can only estimate the critical
exponent $\eta=\eta_{\bot}$ characterizing the decay of
the transversal spin current correlations.

We have 
studied
zigzag chains of several lengths $L$ ranging from
$64$ to $256$.
It turned out that a relatively large number $m$ of representative states was
necessary to reach good convergence; depending on $S$, $L$ and $S_{\rm tot}$, we
have used $m$ ranging from $400$ to $1500$.
The scalar spin current correlator  $\langle
\vec{\kappa}(n)\cdot \vec{\kappa}(n')\rangle$ has been calculated
for a large number of ground
states in sectors with different total spin $S_{\rm tot}$.

\begin{figure}[tb]
\centerline{\includegraphics[width=0.65\textwidth]{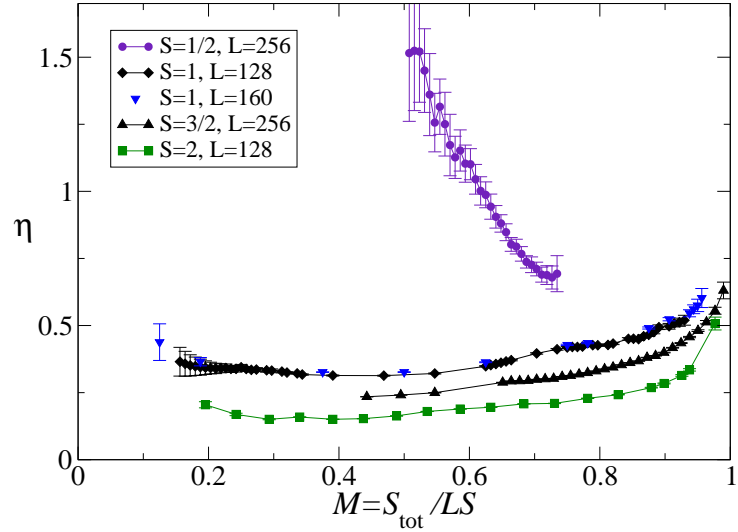}}
\caption{Critical exponent $\eta$, which determines the power-law decay of the
  transversal spin current correlations, as a function of the magnetization per
  spin $MM=S_{\rm tot}/LS$, for isotropic spin-$S$ chains with the spin $S=1/2$,
  $1$, $3/2$ and $2$. The error bars shown correspond to the uncertainties of
  the fit. }
\label{fig:eta}
\end{figure}

The correlator has been averaged over the starting and final positions $n$,
$n'$, and contributions with $n$ or $n'$ being closer as as a fixed ``cutoff''
(taken here to be $20$ sites) to the chain ends were
discarded.
The DMRG data for the correlator has been fitted to the power-law form
\begin{equation} 
\label{fit-law} 
\langle\vec{\kappa}(x)\cdot \vec{\kappa}(0)\rangle =\kappa_{0}^{2}+ Ax^{-\eta}\cos[q(x+\delta)]
\end{equation}
suggested by (\ref{kappa-corr}), (\ref{kappa-corr-Hs}); the introduction of a
finite phase shift $\delta$ helps to suit the open boundary
conditions. From those fits we have extracted the behavior of the equilibrium
spin current $\kappa_{0}$ and the exponent $\eta$ as functions of the chain
magnetization $M=S_{\rm tot}/L$, shown respectively in Fig.\ \ref{fig:kappa} and
Fig.\ \ref{fig:eta}. The wave vector $q$ extracted from the fit depends only
weakly on the magnetization, for all $S$ studied, so we do not present the
corresponding dependences. 

The quality of fits is generally at its best in the
intermediate $M$ region, and is deteriorating at $M\to0$ or $M\to1$ for the
reasons discussed in Ref.\ \cite{McCulloch+08}. For small $M$ it was only
possible to extract the asymptotic value $\kappa_{0}^{2}$ of the spin current
correlator, but not the critical exponent $\eta$.  The extracted value of $\eta$
is weakly depending on $M$ and for $S\geq 1$ qualitatively agrees with the theoretical
estimates predicting that $\eta$ should vary from approximately
$1/4$ to $1/2$ as $M$ varies from $0$ to $1$. One should realize that the error
bars shown in Figs.\ \ref{fig:kappa},\ref{fig:eta} are only of indicative nature,
since they show just the uncertainties of the fit to the fixed form
(\ref{fit-law}) and do not take into account the variations of the fit
parameters which would result from adding to it subleading contributions.

\section{Summary and discussion}
\label{sec:summary}

We have studied spin-$S$ isotropic antiferromagnetic frustrated chains in strong
magnetic fields, described by the model (\ref{S-ham}) with the frustration parameter $\alpha=1$ and the spin $S$
ranging from $1/2$ to $2$, by means of the matrix product density matrix
renormalization group technique.  Existence of a phase with field-induced
spontaneous spin current (vector chiral, or $p$-type nematic order) is
established for all $S$ studied. For $S\geq 1$, the behavior of the order
parameter and its correlations as functions of the magnetization qualitatively
agrees with the theory proposed in Ref.\ \cite{KV05}. For $S=1/2$, the existence
of the vector chiral phase and the transition between the chiral and XY2 phases
are captured well by the large-$\alpha$ bosonization approach of
Ref.\ \cite{KV05}.  

However, the behavior of $S=1/2$ chain close to the
saturation $M=1$  does not fit into the common large-$S$ scheme of \cite{KV05}: there is a
finite non-chiral region in the immediate vicinity of $M=1$ which, as shown in
\cite{Okunishi08}, belongs to the two-component Tomonaga-Luttinger liquid (TLL)
phase. This peculiarity might be attributed to the fact that the large-$S$
approach in \cite{KV05} is based on identifying the two emerging particle
species as bosons, while the underlying particles for $S=1/2$ are the Jordan-Wigner
fermions. From the Bethe-ansatz results for the Hubbard model
\cite{1DHubbard-book} it is known that pure density-density interaction does not
destroy the two-component TLL in the low particle density limit. Thus, for
$S=1/2$ the nature of
the transition from the two-component TLL into the chiral phase close to the
saturation field still remains to be explained.

\section*{Acknowledgements}
We are grateful to T. Vekua for fruitful discussions.
AK was supported by the Heisenberg Program (grant KO~2335/1-2) of Deutsche
Forschungsgemeinschaft.

%
%
  \label{last@page}

\begin{thebibliography}{10}

\bibitem{Villain78} J. Villain, Ann. Isr. Phys. Soc. \textbf{2}, 565 (1978).


\bibitem{AndreevGrishchuk84} A.F.Andreev and I.A. Grishchuk, Sov. Phys. JETP 60, 267 (1984).

\bibitem{Kaburagi+99} M. Kaburagi, H. Kawamura, and T. Hikihara,
 J. Phys. Soc. Jpn. \textbf{68}, 3185 (1999).
 
\bibitem{Hikihara+01} T. Hikihara, M. Kaburagi, and H. Kawamura, 
Phys. Rev. B {\bf 63}, 174430 (2001).

\bibitem{Shannon+06}
N. Shannon, T. Momoi, P. Sindzingre, Phys. Rev. Lett. 96, 027213 (2006).

\bibitem{McCulloch+08} I. P. McCulloch, R. Kube, M. Kurz, A. Kleine, U. Schollw\"ock, A. K. Kolezhuk, 
   Phys. Rev. B \textbf{77}, 094404 (2008).

\bibitem{Hikihara+08} T. Hikihara, L. Kecke, T. Momoi, A. Furusaki, Phys. Rev. B
  \textbf{78}, 144404 (2008);
J. Sudan, A. Luscher, A. Laeuchli, preprint arXiv:0807.1923. 


\bibitem{ChandraColeman91} P. Chandra and P. Coleman,
  Phys. Rev. Lett. \textbf{66}, 100 (1991).

\bibitem{Nersesyan+98} A. A. Nersesyan, A. O. Gogolin, and
F. H. L. E{\ss}ler, Phys. Rev. Lett. {\bf 81}, 910 (1998).


\bibitem{K00} A. K. Kolezhuk, Phys. Rev. B {\bf 62}, R6057 (2000).

\bibitem{Lecheminant+01} P. Lecheminant, T. Jolicoeur, and P. Azaria, Phys. Rev. B
{\bf 63}, 174426 (2001).

\bibitem{KV05} A. Kolezhuk and T. Vekua,  
Phys. Rev. B \textbf{72}, 094424 (2005).

\bibitem{Okunishi08} K. Okunishi, J. Phys. Soc. Jpn. \textbf{77}, 114004 (2008).

\bibitem{Enderle+05} M. Enderle, C. Mukherjee,  B. Fak,  R. K. Kremer,  J.-M.
	Broto,  H. Rosner,  S.-L. Drechsler,  J. Richter,  J. Malek,
	A. Prokofiev,  W. Assmus,  S. Pujol,  J.-L. Raggazzoni,
	H. Rakoto,  M. Rheinst\"adter, and  H. M. Ronnow,
        Europhys. Lett. \textbf{70}, 237 (2005).

\bibitem{Drechsler+07} S.-L. Drechsler, O. Volkova, A. N. Vasiliev, N. Tristan,
 J. Richter, M. Schmitt, H. Rosner, J. M\'{a}lek,  R. Klingeler,  A. A. Zvyagin,
 and B. B\"{u}chner, Phys. Rev. Lett. \textbf{98}, 077202 (2007).


\bibitem{Garlea+09} V. O. Garlea, A. Zheludev, K. Habicht,
  M. Meissner, B. Grenier, L.-P. Regnault, and E. Ressouche,
  Phys. Rev. B \textbf{79}, 060404(R) (2009). 

\bibitem{Bogoliubov+86} N. M. Bogoliubov, A. G. Izergin, and V. E. Korepin,
Nucl. Phys. B {\bf 275}, 687 (1986).

\bibitem{AffleckOshikawa99} I. Affleck and M. Oshikawa, Phys. Rev. B {\bf 60},
1038 (1999).

\bibitem{Furusaki+03-04} 
F. H. L. Essler, A. Furusaki, and T. Hikihara, Phys. Rev. B {\bf 68},
064410 (2003);
T. Hikihara and A. Furusaki, \emph{ibid.} {\bf 69}, 064427 (2004).

\bibitem{KonikFendley02} R. M. Konik and P. Fendley Phys. Rev. B {\bf 66},
  144416 (2002).

\bibitem{CamposVenuti+02} L. Campos Venuti, E. Ercolessi, G. Morandi, 
P. Pieri, and M. Roncaglia, Int. J. Mod. Phys. B {\bf 16}, 1363 (2002).

\bibitem{Sato06} Masahiro Sato, J. Stat. Mech. P09001 (2006).

\bibitem{Fath03} G. Fath, Phys. Rev. B {\bf 68}, 134445 (2003).

\bibitem{Schulz86} H. J. Schulz: Phys. Rev. B \textbf{34}, 6372 (1986).

\bibitem{Vekua+07} T. Vekua, A. Honecker, H.-J. Mikeska, and F. Heidrich-Meisner, 
Phys. Rev. B \textbf{76}, 174420 (2007).

\bibitem{White92} S. R. White, Phys. Rev. Lett. \textbf{69}, 2863 (1992).

\bibitem{Schollwock-RMP-05} U. Schollw\"ock, Rev. Mod. Phys. \textbf{77}, 259 (2005).


\bibitem{McCulloch07} I. P. McCulloch, J. Stat. Mech.: Theor. Exp., P10014 (2007).

\bibitem{McCullochGulacsi02} I. P. McCulloch and M. Gulacsi,
  Europhys. Lett. \textbf{57}, 852 (2002).

\bibitem{1DHubbard-book} F. H. L. Essler,  H. Frahm,  F. G\"ohmann,  A. Kl\"umper, and
	V. E. Korepin, \textit{``The One-Dimensional Hubbard Model''} (Cambridge
        University Press, 2005).

\end{thebibliography}
  \end{document}